\documentclass[10pt,journal,cspaper,compsoc]{IEEEtran}
\usepackage[latin9]{inputenc}
\usepackage{array}
\usepackage{multirow}
\usepackage{amsmath}
\usepackage{amssymb}
\usepackage{graphicx}
\usepackage{subfig}

\begin{document}
\title{Novel Repair-by-Transfer Codes and Systematic Exact-MBR Codes with Lower Complexities and Smaller Field Sizes}
\author{Sian-Jheng Lin and Wei-Ho Chung{*}%
\thanks{Authors are with the Research Center for Information Technology Innovation, Academia Sinica, Taipei City, Taiwan. (e-mail: sjlin@citi.sinica.edu.tw; whc@citi.sinica.edu.tw)%
}}

\IEEEcompsoctitleabstractindextext{
\begin{abstract}
The $(n,k,d)$ regenerating code is a class of $(n,k)$ erasure codes with the capability to recover a lost code fragment from other $d$ existing code fragments. This paper concentrates on the design of exact regenerating codes at Minimum Bandwidth Regenerating (MBR) points. For $d=n-1$, a class of $(n,k,d=n-1)$ Exact-MBR codes, termed as repair-by-transfer codes, have been developed in prior work to avoid arithmetic operations in node repairing process. The first result of this paper presents a new class of repair-by-transfer codes via congruent transformations. As compared with the prior work, the advantages of the proposed codes include: i). The minimum of the finite field size is significantly reduced from $n \choose 2$ to $n$. ii). The encoding complexity is decreased from $n^4$ to $n^3$. As shown in simulations, the proposed repair-by-transfer codes have lower computational overhead when $n$ is greater than a specific constant. The second result of this paper presents a new form of coding matrix for product-matrix Exact-MBR codes. The proposed coding matrix includes a number of advantages: i). The minimum of the finite field size is reduced from $n-k+d$ to $n$. ii). The fast Reed-Solomon erasure coding algorithms can be applied on the Exact-MBR codes to reduce the time complexities.
\end{abstract}
}

\maketitle

\section{Introduction}
\IEEEPARstart{I}n a distributed storage system, the source data (message) is dispersed across nodes in the network, and a data collector (DC) can retrieve the whole source data by accessing a subset of the nodes. To tradeoff between the storage efficiency and the system reliability, the erasure codes, such as maximum-distance-separable (MDS) codes \cite{Reed}, random linear codes \cite{Lin} or fountain codes \cite{Luby,Shokrollahi}, are usually adopted as the base of data format in distributed storage systems \cite{Xia,Rhea,Khan}. For an unstable network, the nodes may frequently join and depart. When a node departs or crashes, the system manager will place a replacement node in the distributed storage network to replace the functionality of the failed node. Suppose the replacement node does not store any information about the data (code fragments) stored in the prior failed node. To reconstruct the data, the replacement node broadcasts a request to a subset of other helper nodes, and those helper nodes reply with the requisite information to the replacement node. If the distributed storage systems is based on conventional Reed-Solomon (RS) codes, an intuitive method is to reconstruct the entire source data in the replacement node, and then extract the desired code fragment from the source data. By such method, the total amount of downloaded symbols is not less than the size of whole source data. However, as the size of data stored in a single node is much smaller than the entire source data, it is possible to design a new class of storage codes to reduce the amount of downloaded symbols in node-repairing process. The new class of storage codes, termed as regenerating codes, is introduced by the pioneer paper \cite{Dimakis}. 

\subsection{Coding system description}
In this paper, the regenerating code over $GF(q)$ is associated with a set of parameters $\{n,k,d,\alpha,\beta,B\}$ elaborated in the following. The value $B$ is the number of source symbols over $GF(q)$ to be encoded. The $n$ is the number of produced code fragments, which will be respectively stored in $n$ network nodes. The $\alpha$ is the number of symbols of a code fragment. In data reconstruction process, the DC individually downloads $\alpha$ symbols from each of a subset of $k$ nodes to reconstruct the message. In the node-repairing process, the replacement node individually downloads $\beta$ symbols from each of a subset of $d$ integrity nodes to rebuild the code fragment. Those parameters $\{n,k,d\}$ follows the inequality 
\[
k\leq d\leq n-1.
\]
The theoretical bound of storage-bandwidth trade-off have been given by \cite{Wu} based on the cut-set bound of network coding:
\begin{equation}
B\leq\sum_{i=0}^{k-1}min\{\alpha,(d-i)\beta\}.\label{eq:B}
\end{equation}
By the theoretical bound (\ref{eq:B}), two extreme points on the storage-bandwidth trade-off have been adequately investigated in prior works. The first extreme point, termed as minimum storage regeneration (MSR) point, is firstly to minimize the $\alpha$ and then minimize the $\beta$. The parameter configuration is
\begin{equation}
\begin{array}{l}
\alpha=B/k;\\
\beta=B/\left(k(d-k+1)\right).
\end{array}\label{eq:parameter_MSR}
\end{equation}
The second extreme point, termed as the minimum bandwidth regenerating (MBR) point, is firstly to minimize the $\beta$, and then minimize the $\alpha$. The parameter configuration is
\begin{equation}
\begin{array}{l}
\beta=2B/\left(k(2d-k+1)\right);\\
\alpha=d\beta.
\end{array}\label{eq:parameter_MBR}
\end{equation}
By the so-called data striping technique \cite{Rashmi}, the regenerating codes at $\beta=1$ can be used to construct the regenerating codes for any $\beta$. Thus, here in after, we focus on the design of regenerating codes at the $\beta=1$ MBR points, and the corresponding parameter configuration is
\begin{equation}
\alpha=d,\beta=1\textrm{, and }B={k+1 \choose 2}+k(d-k).\label{eq:Exact-MBR-parameter}
\end{equation}
In the node-regenerating process, if the restored fragment is always the same with the fragment in the prior failed node, this property is called the exact regeneration. This is in contrast to the functional regeneration without imposing restrictions on the content of the stored fragment. Practically, the exact regeneration is a good property to simplify the hardware and software designs for distributed storage systems. However, the non-existence of exact regeneration codes at the interior points on the storage-bandwidth trade-off curve have been proved \cite{Shah}. In this paper, the abbreviations "Exact-MSR" and "Exact-MBR" respectively indicate the regenerating codes at MSR and MBR points with the exact regeneration property.

\subsection{Definitions of terminologies}
\subsubsection{Systematic regenerating codes}
The \cite{Rashmi} defines the systematic regenerating code as a class of regenerating code whose $B$ message symbols appear on a certain set of $k$ systematic code fragments. The nodes storing those systematic fragments are termed as the systematic nodes. A major work of this paper is to construct the systematic regenerating codes at MBR points. Systematic codes are useful in data reconstruction: If the DC can download those systematic code fragments, the DC can directly obtain the corresponding pieces of source data without any computational cost. This is a good property for practical systems.

\subsubsection{Repair-by-transfer codes}
In the node-repairing process, the replacement node broadcasts a request to a subset of helper nodes, and each helper node returns certain number of responding symbols to the replacement node. In general, each helper node should compute the responding symbols via a function of the fragment stored in the node. The repair-by-transfer codes are a class of distributed storage codes where each helper node simply needs to pass a portion of the stored fragment without any arithmetic operations. The repair-by-transfer codes are particularly beneficial to the unstable network environment with frequent occurrence of the node regenerations. A repair-by-transfer code at $(n,k,d=n-1)$ Exact-MBR case is proposed by Shah et al. \cite{Shah}, and the non-existence of other cases $d<n-1$ is shown in \cite{Shah-2}. The details \cite{Shah} are introduced in Section \ref{sub:Comparisons-for-RBT}. Furthermore, the generalized form of \cite{Shah} is presented in \cite{Rouayheb,Pawar}. A system implementation for $k=n-1$ and $k=n-2$ is demonstrated by Hu et al. \cite{Hu-2}. A objective of this paper is to construct the $(n,k,d=n-1)$ Repair-by-transfer codes with smaller finite fields and lower computational costs. By assigning $d=n-1$ to the (\ref{eq:Exact-MBR-parameter}), the parameters for $(n,k,d=n-1)$ repair-by-transfer codes are
\begin{equation}
\alpha=d=n-1,\beta=1\textrm{, and }B=(n-1)k-{k \choose 2}.\label{eq:rbt-parameter}
\end{equation}

\subsubsection{Partial downloading scheme}
By the MBR data reconstruction process in \cite{Rashmi}, the DC should download the whole data stored in the set of connected nodes. To reduce the total amount of downloaded symbols, Gong and Wang \cite{Gong} present a data decoding algorithm, termed as partial downloading scheme, on the non-systematic Exact-MBR codes \cite{Rashmi}. By the partial downloading scheme, the DC can download a partial portion of code fragment from each connected node. The partial downloading scheme is useful to mitigate the network congestion. Thus, the partial downloading schemes are also developed on the proposed repair-by-transfer codes and Exact-MBR codes.

\subsection{Previous works}
The exact regenerating codes at MSR and MBR points have been proposed in recent years. For Exact-MSR codes, the \cite{Cullina} discovers the code constructions at $(n=4,k=2,d=3)$ and ($n=5,k=3,d=4)$ via computer searching. The \cite{Shah-1} presents the Exact-MSR codes for $d=n-1\geq2k-1$ based on interference alignment technique. The non-existence of Exact-MSR code for $d<2k-3$ with $\beta=1$ is shown in \cite{Shah-1}. The \cite{Cadambe,Suh-1} have shown the existence of exact-MSR codes for all $(n,k,d)$, while the size of message approaches infinity. By interference alignment technique, the \cite{Suh} describes the Exact-MSR codes for the following cases: i) $k/n\leq1/2,d\geq2k-1$; and ii) $k\leq3$. Rashmi et al. \cite{Rashmi} present an construction for $(n,k,d\geq2k-2)$ Exact-MSR codes via a product matrix framework. In Exact-MBR codes, the \cite{Shah} presents the $(n=d+1,k,d)$ Exact-MBR codes with no arithmetic operations in node regeneration process, and the \cite{Rashmi} presents the constructions for all feasible $(n,k,d)$ Exact-MBR codes. Furthermore, the cooperative repair codes \cite{Hu,Shum} are the generalized version of regenerating codes to address multiple node failures.

\subsection{Results and organizations of the paper}
In this paper, we developed two classes of Exact-MBR codes. The first result is the repair-by-transfer code at $(n,k,d=n-1)$ Exact-MBR points via the congruences of skew-symmetric matrices. The systematic version and the partial downloading scheme are also proposed.  The second result is the systematic version of Exact-MBR code for all feasible values of $(n,k,d)$ based on the framework defined by \cite{Rashmi}. We design a new encoding matrix for systematic Exact-MBR code, and the partial downloading scheme are also proposed. To emphasize the contributions of the paper, Section \ref{sec:Comparisons-and-discussions} shows the comparisons of the proposed codes with the previous works.

Notations and conventions are declared as follows. Throughout this paper, the operations and symbols are drawn from the field $GF(q)$. For a vector $x$, the underlined notation as $\underline{x}$ represents a row vector, and the over-lined notation as $\overline{x}$ represents a column vector. The $x[i]$ denotes the $i$-th element of the vector $x$. For a matrix $X$, the $X[i,j]$ denotes the entry at $i$-th row and $j$-th column. For a matrix (vector) $X$, the superscript $'t'$ on a matrix (vector) $X^{t}$ denotes the transpose of this $X$. The $I_{k}$ represents a $k\times k$ identity matrix.

The rest of this paper is organized as follows. Section \ref{sec:existing-works} reviews the previous works, such as repair-by-transfer codes and Exact-MBR codes. Section \ref{sec:repair-by-transfer codes} presents the new class of repair-by-transfer codes. 
Section \ref{sec:Systematic-Exact-MBR} presents the proposed systematic Exact-MBR codes based on partially systematic Reed-Solomon (PSRS) codes. Another construction approach is placed in Appendix. The comparisons and discussions are placed in Section \ref{sec:Comparisons-and-discussions}. Section \ref{sec:Conclusions} concludes this paper. 

\section{Previous works\label{sec:existing-works}}
This section reviews a number of related works, such as repair-by transfer codes \cite{Shah}, Exact-MBR codes \cite{Rashmi}, and partial downloading scheme \cite{Gong}.

\subsection{Repair-by-transfer codes \cite{Shah} \label{sec:existing-repair-by-transfer}}
\begin{figure}
\includegraphics[width=1\columnwidth]{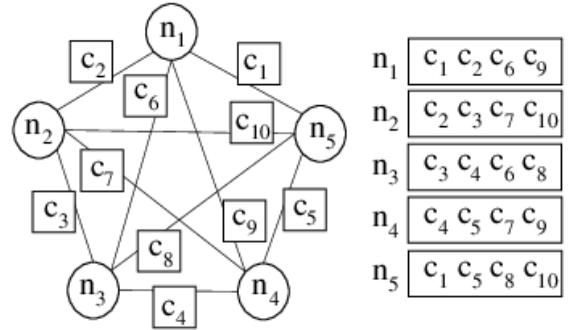}\caption{\label{fig:Shah-rbt-example}Graphical representation of the $(n=5,k=3,d=4)$
repair-by-transfer code proposed by \cite{Shah}.}
\end{figure}

This subsection briefly introduces the $(n,k,d=n-1)$ repair-by-transfer codes \cite{Shah} by a simple example $(n=5,k=3,d=4)$ shown in Figure \ref{fig:Shah-rbt-example}. In beginning, the $B=9$ message symbols are encoded with $(N,K)=\left({n \choose 2},B\right)$ MDS codes, to generate ${n \choose 2}$ code packets. Then each code packet is stored in two distinct nodes. The assignment rule can be visualized with a complete graph of $n$ vertices. As shown in Figure \ref{fig:Shah-rbt-example}, each vertex is recognized as an individual node, and each edge corresponds to a distinct code packet. Each node (vertex) stores the $n-1$ code packets linked to this node. The node regeneration is very simple. If one node fails, the lost $n-1$ code packets in this node can be directly downloaded from each of other $n-1$ nodes. To reconstruct the data, a DC download the code packets from $k$ nodes. It can be shown that the DC accesses a total of $B$ distinct code packets, so the message symbols can be reconstructed via the $\left({n \choose 2},B\right)$ MDS decoding. The \cite{Shah} suggests that the doubly extended RS codes can be chosen as the $(N,K)$ MDS coding technique, and the minimal field size is ${n \choose 2}\leq N+1$. 

\subsection{Exact-MBR codes \cite{Rashmi} and partial downloading scheme \cite{Gong}\label{sec:Exact-MBR-codes}}
This section reviews the Exact-MBR codes \cite{Rashmi} at (\ref{eq:Exact-MBR-parameter}) through product-matrix framework. In code constructions, the $B$ message symbols are formed as a $d\times d$ message matrix $M$, which is then multiplied by an $n\times d$ encoding matrix $\Psi$, resulting in an $n\times d$ 
\begin{equation}
C=\Psi M.\label{eq:encoding}
\end{equation}
code matrix. Let $\underline{c}_i^t$ denote the $i$-th row of $C$, for $1\leq i\leq n$. The $\underline{c}_i^t$ is computed through
\begin{equation}
\underline{c}_{i}^{t}=\underline{\psi}_{i}^{t}M,\label{eq:encoding-row}
\end{equation}
where the $\underline{\psi}_{i}^{t}$ denotes the $i$-th row of $\Psi$. Each $\underline{c}_i^t$ is then stored in a network node with index $i$.

The message matrix $M$ is expressed as
\begin{equation}
M=\begin{bmatrix}S & T\\
T^{t} & \mathbf{0}
\end{bmatrix},\label{eq:message matrix}
\end{equation}
where the $\mathbf{0}$ denotes a $(d-k)\times(d-k)$ zero matrix, the $T$ is a $k\times(d-k)$ matrix filled with $k(d-k)$ distinct message symbols, and the $S$ is a $k\times k$ symmetric matrix determined by ${k+1 \choose 2}$ message symbols. The upper triangular part of $S$ is filled with the message symbols, and other entries assign the corresponding values such that the symmetry holds. Then, the encoding matrix 
\begin{equation}
\Psi=\begin{bmatrix}\Phi & \Delta\end{bmatrix}\label{eq:coding matrix}
\end{equation}
is the concatenation of a $n\times k$ matrix $\Phi$ with a $n\times(d-k)$ matrix $\Delta$. The coding matrix is chosen in such a way that:\\
i) Any \textit{$d$} rows of $\Psi$ are linearly independent;\\
ii) Any \textit{$k$} rows of $\Phi$ are linearly independent.\\
For the non-systematic case, a feasible form of $\Psi$ is a Vandermonde matrix \cite{Rashmi}.

\subsubsection{Node-repairing process}
Suppose the node $f$ fails, and a replacement node is placed in the network to replace the functionality of the failure node. To reconstruct the code fragment (\ref{eq:encoding-row}) in the failure node, the replacement node connects to a subset of $d$ helper nodes $\{h_{1},h_{2},\ldots ,h_{d}\}$. Then each helper node $h_{j}$ computes the scalar value
\begin{equation}
\upsilon_{h_{j}}=\underline{c}_{h_{j}}^{t}\underline{\psi}_{f}, \label{eq:upsilon_h_j}
\end{equation}
and passes this value on to the replacement node. Thus, the replacement node gather $d$ downloaded symbols expressed as a $d$-element column vector
$\Upsilon_{\mathrm{repair}}=[\upsilon_{h_{1}},\upsilon_{h_{2}},\ldots ,\upsilon_{h_{d}}]^{t}$.
By definition, the $\Upsilon_{\mathrm{repair}}$ possesses the equality
\begin{equation}
\Upsilon_{\mathrm{repair}}=C_{\mathrm{repair}}\underline{\psi}_{f}=\Psi_{\mathrm{repair}}M\underline{\psi}_{f},\label{eq:Upsilon_repair}
\end{equation}
where the $C_{\mathrm{repair}}$ is a $d\times\alpha$ matrix consisting of $d$ rows $\{\underline{c}_{h_{1}}^{t},\underline{c}_{h_{2}}^{t},\ldots ,\underline{c}_{h_{d}}^{t}\}$ taken from the $C$, and the $\Psi_{\mathrm{repair}}$ is a $d\times d$ matrix consisting of $d$ corresponding encoding rows $\{\underline{\psi}_{h_{1}}^{t} \underline{\psi}_{h_{2}}^{t},\ldots ,\underline{\psi}_{h_{d}}^{t}\}$. As the $\Psi_{\mathrm{repair}}$ is invertible by the first condition of the MBR encoding matrix, the decoding formula is formulated as
\begin{equation}
\Psi_{\mathrm{repair}}^{-1}\times\Upsilon_{\mathrm{repair}}=M\underline{\psi}_{f}=\underline{c}_{f},\label{eq: c_h_0^t}
\end{equation}
which is the transpose of the desired fragment $\underline{c}_{f}^{t}$.

\subsubsection{Data reconstruction process}
To reconstruct the message, the DC connects to $k$ active nodes $\{i_{1},i_{2},\ldots , i_{k}\}$ and then downloads $\{\underline{c}_{i_{1}}^{t},\underline{c}_{i_{2}}^{t},\ldots ,\underline{c}_{i_{k}}^{t}\}$ from those connected nodes. The $k$ rows $\{\underline{c}_{i_{1}}^{t},\ldots ,\underline{c}_{i_{k}}^{t}\}$ are formulated as a $k\times\alpha$ matrix $C_{\mathrm{DC}}$ following the order $[g_{1},\ldots , g_{k}]$. That is, each $\underline{c}_{i_{j}}^{t}$ is placed at the $g_{j}$-th row of the matrix $C_{\mathrm{DC}}$. In many cases, the sequence $[g_{1},\ldots ,g_{k}]$ can be defined as a monotonically increasing sequence $g_{i}=i$, $1\leq i\leq k$. However, the proposed partial decoding scheme, addressed in Sec. \ref{sub:DataRecAlg-Min}, requires that systematic codeword fragments should be placed at a specific row of $C_{\mathrm{DC}}$. Based on above definitions, the DC accesses $k$ vectors expressed as 
\begin{equation}
C_{\mathrm{DC}}=\Psi_{\mathrm{DC}}M,\label{eq:C_DC}
\end{equation}
where the $\Psi_{\mathrm{DC}}$ denotes a $k\times d$ matrix consisting of $k$ corresponding encoding rows $\{\underline{\psi}_{i_{1}}^{t},\underline{\psi}_{i_{2}}^{t},\ldots ,\underline{\psi}_{i_{k}}^{t}\}$. By definition (\ref{eq:coding matrix}), the $k\times d$ matrix $\Psi_{\mathrm{DC}}$ can be represented as the concatenation of two sub-matrices, given by
\begin{equation}
\Psi_{\mathrm{DC}}=\begin{bmatrix}\Phi_{\mathrm{DC}} & \Delta_{\mathrm{DC}}\end{bmatrix},\label{eq:Psi_DC}
\end{equation}
where the $k\times k$ matrix $\Phi_{\mathrm{DC}}$ and the $k\times(d-k)$ matrix $\Delta_{\mathrm{DC}}$ are drawn from the sub-matrices of $\Phi$ and $\Delta$. Then the (\ref{eq:C_DC}) can be rewritten as 
\begin{equation}
C_{\mathrm{DC}}=\begin{bmatrix}\Phi_{\mathrm{DC}}S+\Delta_{\mathrm{DC}}T^{t} & \Phi_{\mathrm{DC}}T\end{bmatrix}.\label{eq:data_reconstruction_MBR_C_DC}
\end{equation}
The $C_{\mathrm{DC}}$ is split into two parts $C_{\mathrm{DC}}=[\begin{array}{cc} C_{\mathrm{DC}}^{\Phi} & C_{\mathrm{DC}}^{\Delta}\end{array}]$, where the $k$-column part $C_{\mathrm{DC}}^{\Phi}$ corresponds to $\Phi_{\mathrm{DC}}S+\Delta_{\mathrm{DC}}T^{t}$, and the $(d-k)$-column part $C_{\mathrm{DC}}^{\Delta}$ corresponds to $\Phi_{\mathrm{DC}}T$. Then the (\ref{eq:data_reconstruction_MBR_C_DC}) is reformulated as 
\begin{equation}
C_{\mathrm{DC}}^{\Phi}=\Phi_{\mathrm{DC}}S+\Delta_{\mathrm{DC}}T^{t};\label{eq:C_DC_1}
\end{equation}
\begin{equation}
C_{\mathrm{DC}}^{\Delta}=\Phi_{\mathrm{DC}}T.\label{eq:C_DC_2}
\end{equation}
As the $\Phi_{\mathrm{DC}}$ is non-singular by the second condition of the encoding matrix, the DC can compute the matrix $T=\Phi_{\mathrm{DC}}^{-1}C_{\mathrm{DC}}^{\Delta}$, and subsequently, the $S=\Phi_{\mathrm{DC}}^{-1}(C_{\mathrm{DC}}^{\Phi}-\Delta_{\mathrm{DC}}T^{t})$.

\subsubsection{Partial downloading scheme}
Chen and Wang \cite{Gong} indicate that the above data reconstruction process involves a certain amount of redundancy. In the data reconstruction process, the DC completely downloads $k$ vectors $\{\underline{c}_{i_{j}}^{t}|j=1,\ldots , k\}$ with length $d$ for each $\underline{c}_{i_{j}}^{t}$, to be used to reconstruct the $B={k+1 \choose 2}+k(d-k)$ message symbols. As $(kd-B)={k \choose 2}\geq0$, this process potentially downloads ${k \choose 2}$ redundant symbols. To avoid the wasted transmission resource, the \cite{Gong} develops a partial downloading scheme on the Exact-MBR code. By the scheme, the DC can only download the $C_\mathrm{DC}^\Delta$ and the upper triangular part of $C_\mathrm{DC}^\Phi$. Totally, the DC exactly download $B$ symbols.

In data reconstruction process, the sub-matrix $T$ can be solved by the equality (\ref{eq:C_DC_2}). Let
\begin{equation}
D_{\mathrm{DC}}=C_{\mathrm{DC}}^{\Phi}-\Delta_{\mathrm{DC}}T^{t}\label{eq:D_DC-2}
\end{equation}
denote the solvable part in (\ref{eq:C_DC_1}). Thus, the (\ref{eq:C_DC_1}) is rewritten as 
\begin{equation}
\Phi_{\mathrm{DC}}S=D_{\mathrm{DC}}.\label{eq:D_DC-1}
\end{equation}
In the scheme \cite{Gong}, the DC only downloads the upper triangular part of $C_\mathrm{DC}^\Phi$, so the upper triangular part of $D_{\mathrm{DC}}$ is also accessible. The main idea of solving (\ref{eq:D_DC-1}) is to utilize the symmetry of $S$. The process can be divided into $k$ stages, and each stage solves a column of $S$ in the backward order. While the $d$-th column of $S$ have been solved, the $d$-th row of $S$ is also obtained by symmetry of $S$. The obtained $d$-th row of $S$ will be utilized in the later decoding stages. By such recursive decoding process, a symmetric matrix $S$ can be completely solved.

\section{Repair-by-transfer codes \label{sec:repair-by-transfer codes}}
This section proposes a new class of $(n,k,d=n-1)$ repair-by-transfer codes at (\ref{eq:rbt-parameter}). Upon describing the code constructions, two basic entities, termed as the message matrix $\hat{M}$ and the encoding matrix $\hat{\Phi}$, are defined as follows. The $\hat{M}$ is a $n\times n$ matrix constructed from two sub-matrices $\hat{S}$ and $\hat{T}$. The $\hat{S}$ is a $k\times k$ skew-symmetric matrix determined by ${k \choose 2}$ message symbols. The skew-symmetric matrix is defined as a square matrix $A$ satisfying $A=-A^{t}$. For each entry $A[i,j]$ in the skew-symmetric matrix, the equality holds $A[i,j]=-A[j,i]$. Note that the diagonal entries of skew-symmetric matrix $A$ are filled with zeros $A[i,i]=0$. By the above definition, the strictly upper triangular part of $\hat{S}$ (excluding the diagonal entries) is filled with ${k \choose 2}$ message symbols, and the lower triangular part $\hat{S}$ is filled with the corresponding values such that the skew symmetric condition holds. The remaining $B-{k \choose 2}=k(n-k)$ message symbols are formed as the second matrix $\hat{T}$ with $k\times(n-k)$. The $n\times n$ message matrix $\hat{M}$ is defined as 
\begin{equation}
\hat{M}=\begin{bmatrix}\hat{S} & \hat{T}\\
-\hat{T}^{t} & \mathbf{0}
\end{bmatrix},\label{eq:rbt-message}
\end{equation}
where the $\mathbf{0}$ denotes a $(n-k)\times(n-k)$ zero matrix. Notably, the  $\hat{M}$ is also a skew-symmetric matrix.

For the encoding matrix, this matrix is defined as a $n\times n$ square matrix of the form 
\begin{equation}
\hat{\Psi}=\begin{bmatrix}\hat{\Phi} & \hat{\Delta}\end{bmatrix},
\end{equation}
where the size of the matrix $\hat{\Phi}$ is $n\times k$, and the size of matrix $\hat{\Delta}$ is $n\times(n-k)$. The $\hat{\Psi}$ is chosen in such a way that \\
i) Any $k$ rows of $\hat{\Phi}$ are linearly independent;\\
ii) The matrix $\hat{\Psi}$ is non-singular.\\
The above conditions can be met by choosing $\hat{\Phi}$ to be a $n\times k$ Vandermonde matrix, and the $\hat{\Delta}$ is defined as
\begin{equation}
\hat{\Delta}=\begin{bmatrix}\mathbf{0}\\
I_{n-k}
\end{bmatrix},\label{eq:hatDelta}
\end{equation}
where the $\mathbf{0}$ is a $k\times(n-k)$ zero matrix, and the $I_{n-k}$ is a $(n-k)\times(n-k)$ identity matrix. By above definitions, the feasible range of $n$ is $n\leq q$ over $GF(q)$. Furthermore, the $\hat{\Phi}$ can adopt the extended Vandermonde matrix, which is the encoding matrix of the $(q+1,k)$ doubly extended RS code, as the form. Then the $n$ can be extended to $n=q+1$.

By above matrices, the construction of repair-by-transform code is formulated as a congruence
\[
\hat{C}=\hat{\Psi}\hat{M}\hat{\Psi}^{t}.
\]
There is a useful theorem used in the code constructions: The $n\times n$ matrix $\hat{C}$ congruent to a skew-symmetric matrix $\hat{M}$ is also skew-symmetric. Next, we modify the $\hat{C}$ to obtain a symmetric one $\check{C}$. Each entry in strictly lower triangular part of $\hat{C}$ is replaced with its negation value, resulting in a symmetric matrix $\check{C}$. Equivalently, for each row $\hat{c}_{j}^{t}$ in $\hat{C}$, a modified row $\check{c}_{j}^{t}$ in $\check{C}$ is obtained by assigning each entry to 
\begin{equation}
\check{c}_{j}^{t}[i]=\begin{cases}
\hat{c}_{j}^{t}[i] & \textrm{if }i\geq j;\\
-\hat{c}_{j}^{t}[i] & \textrm{otherwise.}
\end{cases}\label{eq:c_j^t[i]}
\end{equation}

The output $\check{C}$ is the generated codewords. The $n$ rows of $\check{C}$ are then respectively stored in $n$ distinct nodes. For $1\leq i\leq n$, the $n$-element row $\check{c}_{i}^{t}$ is stored in an individual network node indexed as $i$. As the diagonal entries $\{\check{c}_{i}^{t}[i]=0\}_{i=1}^{n}$ are always zeros, those zero symbols do not require storage space. Thus, each node takes $n-1$ units of memory space to store a row of $\check{C}$, and the parameter configuration (\ref{eq:rbt-parameter}) holds $\alpha=n-1$.

\emph{Example 1:} We give an example for $(n=5,k=3)$ repair-by-transfer codes over $GF(4)$. By (\ref{eq:rbt-parameter}), other parameters are set as $d=\alpha=4$, $\beta=1$, and $B=9$. By the definition of message matrix (\ref{eq:rbt-message}), the matrix $\hat{M}$ is filled with 5 message symbols $\{u_{i}\}_{i=1}^{9}$ as follows: 
\[
\hat{M}=\begin{bmatrix}0 & u_{1} & u_{2} & u_{3} & u_{4}\\
-u_{1} & 0 & u_{5} & u_{6} & u_{7}\\
-u_{2} & -u_{5} & 0 & u_{8} & u_{9}\\
-u_{3} & -u_{6} & -u_{8} & 0 & 0\\
-u_{4} & -u_{7} & -u_{9} & 0 & 0
\end{bmatrix}.
\]
As $-u_{i}=u_{i}$ over the field of characteristic two, the $\hat{M}$ is also a symmetric matrix. For the encoding matrix, the matrix $\hat{\Phi}$ is chosen as the $5\times3$ extended Vandermonde matrix given by 
\[
\hat{\Phi}=\begin{bmatrix}1 & 0 & 0\\
1 & 1 & 1\\
1 & \omega & \omega^{2}\\
1 & \omega^{2} & \omega^{4}\\
0 & 0 & 1
\end{bmatrix},
\]
where the $\omega$ denotes the primitive element of $GF(4)$. By the $\hat{\Delta}$ defined in (\ref{eq:hatDelta}), the encoding matrix is expressed as
\[
\hat{\Psi}=\begin{bmatrix}1 & 0 & 0 & 0 & 0\\
1 & 1 & 1 & 0 & 0\\
1 & \omega & \omega^{2} & 0 & 0\\
1 & \omega^{2} & \omega^{4} & 1 & 0\\
0 & 0 & 1 & 0 & 1
\end{bmatrix}.
\]
As the $\hat{M}$ is skew-symmetric, the congruence $\hat{C}=\hat{\Psi}\hat{M}\hat{\Psi}^{t}$ is also skew-symmetric, expressed as
\begin{equation}
\hat{C}=\begin{bmatrix}0 & c_{1} & c_{2} & c_{3} & c_{4}\\
-c_{1} & 0 & c_{5} & c_{6} & c_{7}\\
-c_{2} & -c_{5} & 0 & c_{8} & c_{9}\\
-c_{3} & -c_{6} & -c_{8} & 0 & c_{10}\\
-c_{4} & -c_{7} & -c_{9} & -c_{10} & 0
\end{bmatrix}.\label{eq:exp-C}
\end{equation}
Then each entry of strictly lower triangular part of $\hat{C}$ is replaced with its additive inverse value, resulting in
\begin{equation}
\check{C}=\begin{bmatrix}0 & c_{1} & c_{2} & c_{3} & c_{4}\\
c_{1} & 0 & c_{5} & c_{6} & c_{7}\\
c_{2} & c_{5} & 0 & c_{8} & c_{9}\\
c_{3} & c_{6} & c_{8} & 0 & c_{10}\\
c_{4} & c_{7} & c_{9} & c_{10} & 0
\end{bmatrix}.\label{eq:exp-C2}
\end{equation}
Notably, as the $\hat{C}$ is over the field of characteristic two, the $\check{C}=\hat{C}$ can be directly obtained without any arithmetic operations.

\subsection{Node-repairing process}
The node-repairing process utilizes the symmetry of $\check{C}$. Suppose the node $h_{0}$ fails, and the failure node stores the vector $\check{c}_{h_{0}}^{t}$ at the $h_{0}$-th row of $\check{C}$. By the symmetry of $\check{C}$, the $h_{0}$-th row of $\check{C}$ is equivalent to the $h_{0}$-th column of $\check{C}$, whose entries (excluding the entry at main diagonal) are respectively stored in $n-1$ non-failure nodes. Thus, the replacement node can directly download the elements at the $h_{0}$-th column $\check{C}$ from other $n-1$ nodes. Let the $\check{c}_{j}^{t}[i]$ denote the $i$-th element of the row $\check{c}_{j}^{t}$. The formulation is given by 
\begin{equation}
\check{c}_{h_{0}}^{t}[i]=\begin{cases}
0 & \textrm{if }i=h_{0};\\
\check{c}_{i}^{t}[h_{0}] & \textrm{Otherwise.}
\end{cases}
\end{equation}
Consequently, this node-repairing process does not involve any arithmetic operations at the helper nodes and the replacement node, as illustrated in Example 1. In (\ref{eq:exp-C2}), if any one row of $\check{C}$ is erased, this row can be regenerated through the aid of corresponding column in $\check{C}$.

\subsection{Data reconstruction process with full downloading \label{sub:DataRec-full-1}}
In data reconstruction, the DC accesses the $k$ rows $\{\check{c}_{i_{1}}^{t},\check{c}_{i_{2}}^{t},\ldots ,\check{c}_{i_{k}}^{t}\}$, which are respectively downloaded from $k$ connected nodes $\{i_{1},i_{2},\ldots , i_{k}\}$. To begin with, each row $\check{c}_{i_{j}}^{t}$, $1\leq j\leq k$, is restored to the original vector $\hat{c}_{i_{j}}^{t}$ via the inversion of formula (\ref{eq:c_j^t[i]}). The restored results $\{\hat{c}_{i_{1}}^{t},\ldots ,\hat{c}_{i_{k}}^{t}\}$ are formed as a $k\times n$ matrix $\hat{C}_{\mathrm{DC}}$ which is a sub-matrix of $\hat{C}$. By construction, the $\hat{C}_{\mathrm{DC}}$ possesses the equality 
\begin{equation}
\hat{C}_{\mathrm{DC}}=\hat{\Psi}_{\mathrm{DC}}\hat{M}\hat{\Psi}^{t},\label{eq:hatC_mathrmDC}
\end{equation}
where the $k\times(n-1)$ matrix 
\[
\hat{\Psi}_{\mathrm{DC}}=\begin{bmatrix}\hat{\Phi}_{\mathrm{DC}} & \hat{\Delta}_{\mathrm{DC}}\end{bmatrix}=\left[\begin{array}{c}
\psi_{i_{1}}^{t}\\
\vdots\\
\psi_{i_{k}}^{t}
\end{array}\right]
\]
consists of the $k$ encoding rows of $\hat{C}_{\mathrm{DC}}$.

As the $\hat{\Psi}$ is non-singular by the second condition of $\hat{\Psi}$, the $\hat{C}_{\mathrm{DC}}$ in (\ref{eq:hatC_mathrmDC}) is then post-multiplied by its inversion $(\hat{\Psi}^{t})^{-1}$, to obtain a $k\times n$ matrix
\begin{equation}
\hat{D}_{\mathrm{DC}}=\hat{C}_{\mathrm{DC}}(\hat{\Psi}^{t})^{-1}=\hat{\Psi}_{\mathrm{DC}}\hat{M}.\label{eq:D_DC}
\end{equation}
The term $\hat{\Psi}_{\mathrm{DC}}\hat{M}$ in (\ref{eq:D_DC}) is then decomposed as two parts:
\[
\hat{\Psi}_{\mathrm{DC}}\hat{M}=\begin{bmatrix}\hat{\Phi}_{\mathrm{DC}}\hat{S}-\hat{\Delta}_{\mathrm{DC}}\hat{T}^{t} & \hat{\Phi}_{\mathrm{DC}}\hat{T}\end{bmatrix}.
\]
To elaborate the process, the $\hat{D}_{\mathrm{DC}}$ is split into two parts $\hat{D}_{\mathrm{DC}}=[\begin{array}{cc}\hat{D}_{\mathrm{DC}}^{\Phi} & \hat{D}_{\mathrm{DC}}^{\Delta}\end{array}]$, where the left part $\hat{D}_{\mathrm{DC}}^{\Phi}$ has $k$ columns and the right part $\hat{D}_{\mathrm{DC}}^{\Delta}$ has $(n-k)$ columns, so
\begin{equation}
\hat{D}_{\mathrm{DC}}^{\Phi}=\hat{\Phi}_{\mathrm{DC}}\hat{S}-\hat{\Delta}_{\mathrm{DC}}\hat{T}^{t};\label{eq:D_DC_1}
\end{equation}
\begin{equation}
\hat{D}_{\mathrm{DC}}^{\Delta}=\hat{\Phi}_{\mathrm{DC}}\hat{T}.\label{eq:D_DC_2}
\end{equation}
By the first definition of $\hat{\Psi}$, the $\hat{\Phi}_{\mathrm{DC}}$ is non-singular. Thus, the DC can compute the matrix $\hat{T}=\hat{\Phi}_{\mathrm{DC}}^{-1}\hat{D}_{\mathrm{DC}}^{\Delta}$;
and subsequently, the $\hat{S}=\hat{\Phi}_{\mathrm{DC}}^{-1}(\hat{D}_{\mathrm{DC}}^{\Phi}+\hat{\Delta}_{\mathrm{DC}}\hat{T}^{t})$.

\subsection{Systematic version of repair-by-transfer codes\label{sub:Coding-RBT}}
To construct a systematic version of repair-by-transfer codes, a message-symbol remapping procedure is employed to determine the entries of $\hat{M}$. Without loss of generality, we declare that the source data are embedded in the first $k$ rows of $\hat{C}$. To reduce the computational cost, the $\hat{\Phi}$, which is the sub-matrix of $\hat{\Psi}$, is defined as the encoding matrix of $(n,k)$ systematic RS codes. The matrix contains two parts expressed as
\begin{equation}
\hat{\Phi}=\begin{bmatrix}I_{k}\\
\check{\Phi}
\end{bmatrix},\label{eq:hat{Phi}}
\end{equation}
where the first $k$ rows of $\hat{\Phi}$ is an identity matrix $I_{k}$. By the $\hat{\Delta}$ defined in (\ref{eq:hatDelta}), the encoding matrix $\hat{\Psi}$ is thus formulated as 
\[
\hat{\Psi}=\begin{bmatrix}I_{k} & \mathbf{0}\\
\check{\Phi} & I_{n-k}
\end{bmatrix}.
\]
By the above encoding matrix, the encoding formula $\hat{C}=\hat{\Psi}\hat{M}\hat{\Psi}^{t}$ can be rewritten as

\[
\hat{C}=\begin{bmatrix}S & S\check{\Phi}^{t}+T\\
\check{\Phi}S-T^{t} & \check{\Phi}S\check{\Phi}^{t}+\check{\Phi}T-T^{t}\check{\Phi}^{t}
\end{bmatrix}.
\]
To achieve the systematic condition, the first $k$ rows of $\hat{C}$, expressed as $\begin{bmatrix}S & S\check{\Phi}^{t}+T\end{bmatrix}$, are defined as the source data. Let $U=[\begin{array}{cc} U_{\mathrm{L}} & U_{\mathrm{R}}\end{array}]$ denote a $k\times n$ matrix consisting of $B$ source symbols. The $U_{\mathrm{L}}$ is a $k\times k$ skew-symmetric matrix whose strictly upper-triangular part is filled with ${k \choose 2}$ source symbols, and other entries are filled with the corresponding values to satisfy the skew-symmetry condition. The $U_{\mathrm{R}}$ is a $k\times(n-k)$ matrix filled with $k\times(n-k)$ source symbols. The systematic condition gives two equations 
\[
S=U_{\mathrm{L}};S\check{\Phi}^{t}+T=U_{\mathrm{R}}.
\]
By above two equations, the $\hat{C}$ can be rewritten as
\[
\hat{C}=\begin{bmatrix}U_{\mathrm{L}} & U_{\mathrm{R}}\\
-U_{\mathrm{R}}^{t} & V
\end{bmatrix},
\]
where the $V$ is a $(n-k)\times(n-k)$ matrix defined as $V=\check{\Phi}U_{\mathrm{R}}-U_{\mathrm{R}}^{t}\check{\Phi}^{t}-\check{\Phi}U_{\mathrm{L}}\check{\Phi}^{t}$.
As other three parts of $\hat{C}_{\mathrm{DC}}$, namely $U_{\mathrm{L}}$ and $\pm U_{\mathrm{R}}$, are available without the arithmetic computations, the matrix $V$ is the remaining unknown objective to be computed. It is noted that the matrix $V$ is a skew-symmetric matrix, so is the $\hat{C}$. The computation of $T$ involves the matrix product $\check{\Phi}U_{\mathrm{R}}$ and the congruence $\check{\Phi}U_{\mathrm{L}}\check{\Phi}^{t}$, and the term $U_{\mathrm{R}}^{t}\check{\Phi}^{t}$ can be directly obtained via transposing the result $\check{\Phi}U_{\mathrm{R}}$. As the $\check{\Phi}$ identifies the encoding matrix of parity part in the $(n,k)$ systematic RS codes, the product $\check{\Phi}U_{\mathrm{R}}$ denotes the parity parts of RS codes for each column of $U_{\mathrm{R}}$. For the congruence $\check{\Phi}U_{\mathrm{L}}\check{\Phi}^{t}$, the $(n,k)$ systematic RS encoding is applied on each column of $U_{\mathrm{L}}$ to obtain the parity part $\check{\Phi}U_{\mathrm{L}}$. Then the $(n,k)$ systematic RS encoding is applied on each row of $\check{\Phi}U_{\mathrm{L}}$, resulting in the $\check{\Phi}U_{\mathrm{L}}\check{\Phi}^{t}$ at the parity part. By above steps, the product $\check{\Phi}U_{\mathrm{R}}$ requires $O(k(n-k)^{2})$ operations, and the transformation $\check{\Phi}U_{\mathrm{L}}\check{\Phi}^{t}$ requires $O(2k^{2}(n-k))$ operations.

\subsection{Partial downloading scheme}
For the data reconstruction in Sec. \ref{sub:DataRec-full-1}, we suppose that the DC completely downloads the $k$ vectors $\{\check{c}_{i_{j}}^{t}|j=1,\ldots , k\}$, and the length of each vector $\check{c}_{i_{j}}^{t}$ is $n-1$. Thus, the total number of downloaded symbols is $(n-1)k$, which is much larger than the size of message $B=(n-1)k-{k \choose 2}$. By utilizing the symmetry of $\check{C}$, the DC can exactly download $B$ symbols.

For any two distinct codeword vectors $\check{c}_{i_{j}}^{t}$ and $\check{c}_{i_{l}}^{t}$ in $\check{C}$, we have  $\check{c}_{i_{j}}^{t}[i_{l}]=\check{c}_{i_{l}}^{t}[i_{j}]$ by the symmetric property, so the DC can download this symbol only from either the node $i_{j}$ or the node $i_{l}$. Based on this observation, the $k$ connected nodes can avoid the total of ${k \choose 2}$ symbols to be transmitted. An simple transmission strategy is that, the first node $i_{1}$ transmits the whole $n-1$ symbols $\check{c}_{i_{1}}^{t}$ to the DC. Then the second node $i_{2}$ can only transmit $n-2$ symbols of $\check{c}_{i_{2}}^{t}$ to the DC, as the symbol $\check{c}_{i_{2}}^{t}[i_{1}]=\check{c}_{i_{1}}^{t}[i_{2}]$ does not need to be transmitted. Inductively, the connected node $i_{j}$ can only transmit $n-j$ symbols of $\check{c}_{i_{j}}^{t}$ to the DC, for $j=1,\ldots , n$. The above policy is simple, but the data throughputs for each node is imbalanced. Thus, an alternative transmission policy is presented in the following. It is noted that each node can save $(k-1)/2$ symbols of data transmission on average, and this value is achieved for odd $k$ by the proposed transmission policy. For even $k$, as the value $(k-1)/2$ is not an integer, the proposed transmission policy can save $k/2-1$ symbols in each odd-index node, and $k/2$ symbols in each even-index node.

Given any two connected nodes with indices $i_{j}$, $i_{l}$ and $1\leq i,l\leq k$, we define a decision criterion as
\begin{equation}
D(j,l)=\begin{cases}
min\{j,l\} & \textrm{if }j+l\textrm{ is even};\\
max\{j,l\} & \textrm{otherwise.}
\end{cases}\label{eq:decision_criterion}
\end{equation}
As any two distinct nodes $i_{j}$ and $i_{l}$ simultaneously store a common symbol, the $D(j,l)\in\{j,l\}$ returns the index of the chosen node to avoid the transmission of this common symbol. Hence the DC downloads this element from another un-chosen node. Two examples are given in Figure \ref{fig:ex-t-a-f} tabulating the exhaustive outputs of $D(i,j)$ for $k=5$ and $6$. In the case $k=5$, each node omits two symbols in transmission. In the case $k=6$, the nodes $\{g_{1},g_{3},g_{5}\}$ omit two symbols in transmission, and the nodes $\{g_{2},g_{4},g_{6}\}$ omit three symbols in transmission.

The valid of decision criterion (\ref{eq:decision_criterion}) is explained as follows. Given a node indexed by $X$, we consider the output of $D(X,y)$ for $y=1,\ldots , k$. If the (\ref{eq:decision_criterion}) outputs $X=D(X,y)$ for a specific $y$, the node $X$ can omit the transmission of a symbol, and DC will download this symbol from another node $y$. To satisfy the equality $X=D(X,y)$, the range of $y$ are drawn from $y\in\{\ldots , X-1-2i,\ldots , X-1,X+2,\ldots , X+2i,\ldots\}$ and $1\leq y\leq k$. Thus, there are about $k/2$ distinct symbols of $y$, and the condition for bandwidth balance holds.

\begin{figure} \centering
\subfloat[]{
\begin{tabular}{c|c|c|cc}
\multicolumn{1}{c}{} & \multicolumn{1}{c}{5} & \multicolumn{1}{c}{4} & 3  & 2\tabularnewline
\cline{2-5} 
1  & 1  & 4  & \multicolumn{1}{c|}{1} & \multicolumn{1}{c|}{2}\tabularnewline
\cline{2-5} 
2  & 5  & 2  & \multicolumn{1}{c|}{3} & \tabularnewline
\cline{2-4} 
3  & 3  & 4  &  & \tabularnewline
\cline{2-3} 
4  & 5  & \multicolumn{1}{c}{} &  & \tabularnewline
\cline{2-2} 
\end{tabular}
}
\subfloat[]{
\begin{tabular}{c|c|c|c|cc}
\multicolumn{1}{c}{} & \multicolumn{1}{c}{6} & \multicolumn{1}{c}{5} & \multicolumn{1}{c}{4} & 3  & 2\tabularnewline
\cline{2-6} 
1  & 6  & 1  & 4  & \multicolumn{1}{c|}{1} & \multicolumn{1}{c|}{2}\tabularnewline
\cline{2-6} 
2  & 2  & 5  & 2  & \multicolumn{1}{c|}{3} & \tabularnewline
\cline{2-5} 
3  & 6  & 3  & 4  &  & \tabularnewline
\cline{2-4} 
4  & 4  & 5  & \multicolumn{1}{c}{} &  & \tabularnewline
\cline{2-3} 
5  & 6  & \multicolumn{1}{c}{} & \multicolumn{1}{c}{} &  & \tabularnewline
\cline{2-2} 
\end{tabular}
}

\caption{\label{fig:ex-t-a-f} Two examples of the outputs of decision criterion
$D(j,l)$. (a) $k=5$. (b) $k=6$.}
\end{figure}

\section{Systematic Exact-MBR coding algorithm\label{sec:Systematic-Exact-MBR}}
Based on the framework of $(n,k,d)$ Exact-MBR codes \cite{Rashmi} in Sec. \ref{sec:Exact-MBR-codes}, this section presents a systematic form of encoding matrix $\Psi$, where the feasible range of $n$ are $n\leq q$ over $GF(q)$. Then the partial downloading scheme is developed on the proposed Exact-MBR codes. Upon describing the proposed encoding matrix, the encoding (\ref{eq:encoding}) can be divided into $\alpha$ individual columns given by
\begin{equation}
\overline{c}_{i}=\Psi\overline{m}_{i},\label{eq:encoding-column}
\end{equation}
where the $\overline{m}_{i}$ indicates the $i$-th column of $M$, and the result $\overline{c}_{i}$ is the $i$-th column in $C$. The (\ref{eq:encoding-column}) can be rewritten as
\begin{equation}
\overline{c}_{i}=\Psi\overline{m}_{i}=\begin{bmatrix}\Phi & \Delta\end{bmatrix}\left[\begin{array}{c}
\overline{m}_{i}^{a}\\
\overline{m}_{i}^{b}
\end{array}\right]=\Phi\overline{m}_{i}^{a}+\Delta\overline{m}_{i}^{b},
\end{equation}
where the $\overline{m}_{i}^{a}$ denotes the $k$-element vector located in the upper part of the $\overline{m}_{i}$, and the $\overline{m}_{i}^{b}$ denotes the remaining $(d-k)$-elements located in the lower part of the $\overline{m}_{i}$.

By the first condition of Exact-MBR encoding matrix, the $\overline{m}_{i}$ can be reconstructed from arbitrary $d$ elements in $\overline{c}_{i}$. By the second condition, if the term $\Delta\overline{m}_{i}^{b}$ is given, the $\overline{m}_{i}^{a}$ can be reconstructed from arbitrary $k$ elements in $\overline{c}_{i}$. Under above observations, Section \ref{sub:PSRS} presents a class of modified version of Reed-Solomon codes, termed as partially systematic Reed-Solomon (PSRS) codes, to satisfy those conditions. Section \ref{sub:EncAlg} shows that the encoding matrix of the systematic Exact-MBR codes. Section \ref{sub:DataRecAlg-Min} presents the partial downloading scheme.

\subsection{Partially systematic Reed-Solomon codes\label{sub:PSRS}}
We define the partially systematic Reed-Solomon (PSRS) code associated with three parameters $(n,k,d)$ where $k\leq d<n$. The $n$ is the codeword length, the $d$ is the message length, and the $k$ is the length of systematic part. The input is expressed as a $d$-element vector $\underline{c}=\begin{bmatrix}\underline{a} & \underline{b}\end{bmatrix}$, where the sub-vector $\underline{a}=[a_{1}...a_{k}]$ denotes the $k$ systematic symbols, and the sub-vector $\underline{b}=[b_{1}...b_{d-k}]$ denotes the remaining $d-k$ non-systematic symbols. By definition, the systematic part $\underline{a}$ is embedded in the first $k$ elements of the generated codeword. This subsection presents the constructions of $(n,k,d)$ PSRS codes via the polynomial evaluation approach. Let the $G(x)$ denote the coding polynomial constructed from the message $\underline{c}$. The degree of $G(x)$ is $deg(C(x))<d$. The codeword symbols are the evaluations of $C(x)$ at $n$ distinct points: 
\begin{equation}
\{C(x_{1}),C(x_{2}),\ldots , C(x_{n})\}.\label{eq:poly-evaluation}
\end{equation}
As the code is over $GF(q)$, the code suffices for $n\leq q$. By the partially systematic condition, the first $k$ codeword symbols are equivalent to the systematic message symbols. Thus,
\begin{equation}
C(x_{i})=a_{i},\forall i=1,2,\ldots , k.\label{eq:C(a^i)}
\end{equation}
In the following, the $C(x)$ is properly defined to satisfy the partial systematic condition.

The $C(x)$ is defined as the sum of two polynomials
\begin{equation}
C(x)=\Phi(x)+\Delta(x),
\end{equation}
where the polynomial $\Phi(x)$ is constructed from $\underline{a}$, and the $\Delta(x)$ is constructed from $\underline{b}$. The $\Phi(x)$, and $deg(\Phi(x))<k$, is defined as 
\begin{equation} \label{eq:Phi(x)}
\Phi(x)=\sum_{i=1}^{k}a_{i}\prod_{j\neq i}\frac{x-x_{j}}{x_{i}-x_{j}}.
\end{equation}
This follows the form of Lagrange polynomial. Thus, the $\Phi(x)$ possesses the systematic property: 
\[
\Phi(x_{i})=a_{i},\forall i=1,2,\ldots , k.
\]
The polynomial $\Delta(x)$ is defined as the multiplication of two polynomials: 
\begin{equation}
\Delta(x)=\Gamma(x)B(x).
\end{equation}
The polynomial $\Gamma(x)$ has $k$ roots located in the evaluation points of systematic part:
\begin{equation}
\Gamma(x)=\prod_{i=1}^{k}(x-x_{i}).
\end{equation}
The $B(x)$ is constructed from the $(d-k)$-element vector $\underline{b}$. The $B(x)$ can be chosen as the systematic or non-systematic form. For example, a non-systematic form with geometric progression is expressed as 
\begin{equation}
B(x)=\sum_{i=1}^{d-k}b_{i}x^{i-1}.
\end{equation}

By the above definitions, it can be shown that the partial systematic condition (\ref{eq:C(a^i)}) holds:
\begin{equation}
\begin{aligned}C(x_{i}) & =\Phi(x_{i})+\Delta(x_{i})\\
 & =\Phi(x_{i})+0\times B(x_{i})=a_{i},\forall i=1,2,\ldots , k.
\end{aligned}
\end{equation}

In summary, the encoding algorithm includes four major steps listed as follows:\\
i). Compute the coefficients of $\Phi(x)$.\\
ii). Compute the product $\Delta(x)=\Gamma(x)B(x)$, where the coefficients of $\Gamma(x)$ can be computed in advance.\\
iii). Compute the summation $C(x)=\Phi(x)+\Delta(x)$. \\
iv). Evaluate the values $\{C(x_{1}),\ldots , C(x_{n})\}$ to obtain the codeword symbols.\\
If the encoding algorithm is implemented in the native way, the computational complexities of the four steps are $O(k^2)$, $O(k(d-k))$, $O(k)$, and $O(dn)$, respectively. 

To reduce the complexity complexity, we observe that the fast Fourier transforms (FFT) can be utilized to reduce the computational cost in steps (i), (ii) and (iv). The conceptual ideas are addressed below. In step (i), the (\ref{eq:Phi(x)}) can be calculated via fast Lagrange interpolation \cite{Bini} with complexity $O(k \log^2 k)$. Alternatively, the fast Reed-Solomon encoding algorithms can also be used in (\ref{eq:Phi(x)}). If the code is operated on Fermat field $GF(q+1), q\in \{2, 4, 16, 65536\}$, the (\ref{eq:Phi(x)}) can be calculated via inverse fast Fourier transform with complexity $O(k \log k)$ (see \cite{Lin-1} and \cite{Lin-2}). If the code is operated on finite field with characteristic two $GF(q), q\in\{2, 4, 8,\ldots\}$, the \cite{Didier} proposed an coding algorithm with complexity $O(q \log^2 q)$. The step (ii) is a polynomial multiplication. By using FFT, the complexity can be reduced to $O(d \log d)$. In step (iv), the polynomial evaluations can be computed with FFT, and the complexity is $O(n \log n)$.

It is noted that the PSRS codes can also be implemented with generator polynomials. The details are placed in appendix.

\subsubsection{Full erasure decoding from $d$ codeword symbols}
The message vector $\underline{c}$ can be reconstructed from arbitrary $d$ out of $n$ codeword symbols $\{y_{i}=C(z_{i})|1\leq i\leq d\}$. By the subset of codeword symbols, the $C(x)$ is constructed via Lagrange interpolation: 
\begin{equation}
C(x)=\sum_{i=1}^{d}y_{i}\prod_{j\neq i}\frac{x-z_{j}}{z_{i}-z_{j}}.
\end{equation}
The $C(x)$ is then divided by $\Gamma(x)$ to obtain a quotient $B(x)$ and a remainder $\Phi(x)$. The $k$ evaluations $a_{i}=\Phi(x_{i})$, $1\leq i\leq k$, are the systematic part $\underline{a}$, and the coefficients of $B(x)$ are the non-systematic part $\underline{b}$.

\subsubsection{Partial erasure decoding from $k$ codeword symbols}
Suppose the non-systematic part $\underline{b}$ is given. In this case, we shows that the systematic part $\underline{a}$ can be reconstructed from arbitrary $k$ out of $n$ codeword symbols $\{y_{i}=C(z_{i})|1\leq i\leq k\}$. By the given $\underline{b}$, the polynomial $\Delta(x)$ can be constructed. Then the $k$ evaluation values of $\Phi(x)$ are calculated via
\begin{equation}
\Phi(z_{i})=C(z_{i})-\Delta(z_{i}),\forall i=1...k.
\end{equation}
By the $k$ evaluation values of $\Phi(x)$, the $\Phi(x)$ can be interpolated via Lagrange polynomial, and the $\underline{a}$ is the $k$ evaluations $a_{i}=\Phi(x_{i})$.

\subsection{Encoding matrix of proposed Exact-MBR codes\label{sub:EncAlg}}
As the $(n,k,d)$ PSRS codes satisfy the conditions of Exact-MBR codes, the encoding matrix of $(n,k,d)$ PSRS codes can be chosen as the $\Psi$. For the systematic part $\underline{a}$, the  coding polynomial $\Phi(x)$ formulates a generator matrix corresponding to the component $\Phi$ in encoding matrix $\Psi$. By the definition of $\Phi(x)$, the entries of matrix $\Phi$ are 
\begin{equation}
\Phi[l,i]=\prod_{j=1;j\neq i}^{k}\frac{x_{l}-x_{j}}{x_{i}-x_{j}},\textrm{for }i=1,\ldots ,k.
\end{equation}
Consequently, the first $k$ rows of $\Phi$ is a $k\times k$ identity matrix $I_{k}$. For the non-systematic part $\underline{b}$, the coding polynomial $\Delta(x)$ formulates a generator matrix corresponding to the component $\Delta$ in encoding matrix $\Psi$. By the definition of $\Delta(x)$, the entries of matrix $\Delta$ are
\begin{equation}
\Delta[l,i]=x_{l}^{(i-1)}\Gamma(x_{l})\textrm{, for }i=1,\ldots ,d-k.
\end{equation}
As $\Gamma(x_{l})=0$ for $1\leq l\leq k$, the first $k$ rows of $\Delta$ are entirely filled with zeros. Then the encoding matrix $\Psi$ is obtained by combining the $\Phi$ and $\Delta$. Thus, the first $k$ rows of $\Psi$ are in the form $\begin{bmatrix}I_{k} & \mathbf{0}\end{bmatrix}$, so that the corresponding first $k$ rows of the code matrix $C$ are expressed as $[\begin{array}{cc}S & T\end{array}]$. Hence, the proposed Exact-MBR code is systematic. As stated previously, the proposed $\Psi$ satisfies the two conditions of Exact-MBR encoding matrix, which enables the node-repairing algorithm and data reconstruction algorithm addressed in Sec. \ref{sec:Exact-MBR-codes}.

\emph{Example 2:} We give an example for $(n=6,k=3,d=4)$ Exact-MBR codes over $GF(7)$. By (\ref{eq:Exact-MBR-parameter}), other parameters are set as $\alpha=4$, $\beta=1$, and $B=9$. By the definition of message matrix (\ref{eq:message matrix}), the matrices $M$ is filled with 9 message symbols $\{u_{i}\}_{i=1}^{9}$. The $S$, $T$ and $M$ are given by 
\begin{align*}
S & =\begin{bmatrix}u_{1} & u_{2} & u_{3}\\
u_{2} & u_{5} & u_{6}\\
u_{3} & u_{6} & u_{8}
\end{bmatrix},T=\begin{bmatrix}u_{4}\\
u_{7}\\
u_{9}
\end{bmatrix};\\
M & =\begin{bmatrix}u_{1} & u_{2} & u_{3} & u_{4}\\
u_{2} & u_{5} & u_{6} & u_{7}\\
u_{3} & u_{6} & u_{8} & u_{9}\\
u_{4} & u_{7} & u_{9} & 0
\end{bmatrix}.
\end{align*}
The coding polynomial $C(x)$ of $(n=6,k=3,d=4)$ PSRS code is chosen as 
\begin{align*}
\Phi(x)=& a_{1}\times\frac{(x-2)(x-3)}{2}+a_{2}\times\frac{(x-1)(x-3)}{6}\\
 & +a_{3}\times\frac{(x-1)(x-2)}{2};\\
 \Delta(x)=&(x-1)(x-2)(x-3)b_{1}.
\end{align*}
By above definitions, the corresponding matrices $\Phi$ and $\Delta$ are as follows: 
\[
\Phi=\begin{bmatrix}1 & 0 & 0\\
0 & 1 & 0\\
0 & 0 & 1\\
1 & 4 & 3\\
3 & 6 & 6\\
6 & 6 & 3
\end{bmatrix};\Delta=\begin{bmatrix}0\\
0\\
0\\
6\\
3\\
4
\end{bmatrix}.
\]
The encoding matrix $\Psi=\begin{bmatrix}\Phi & \Delta\end{bmatrix}$ is the combination of $\Phi$ and $\Delta$.

\subsection{Partial downloading scheme\label{sub:DataRecAlg-Min}}
This subsection presents the partial downloading scheme on the proposed systematic Exact-MBR codes. Similar to the \cite{Gong}, the proposed scheme only downloads the entire $C_\mathrm{DC}^\Delta$ and the lower (or upper, alternatively) triangular part of $C_\mathrm{DC}^\Phi$. Precisely, each connected node $i_{j}$ passes a portion of the code fragment $c_{i_{j}}^{t}$ in the lower/upper triangular part of $C_{\mathrm{DC}}$. By (\ref{eq:C_DC_2}), the $T$ can be successfully solved. Then the lower/upper triangular part of $D_{\mathrm{DC}}$ can be computed via (\ref{eq:D_DC-2}). The two cases are respectively considered as follows.

\subsubsection{Data collector downloads the lower triangular part of $C_{\mathrm{DC}}^{\Phi}$
\label{sub:lower triangular C_DC}}
In this case, the DC can access the lower triangular part of $D_{\mathrm{DC}}$. The computational structure can be divided into $k$ stages, and the $l$-th stage solves the $l$-th column $\overline{s}_{l}$ of $S$. In the first stage, as the first column of $D_{\mathrm{DC}}$ are fully located in the lower triangular part of $D_{\mathrm{DC}}$, the first column $\overline{s}_{1}$ of $S$ can be solved successfully. By the symmetry of $S$, the first row of $S$ is also obtained $\underline{s}_1=\overline{s}_1^t$. Let $\underline{i}_{l}^{t}$ denote a row vector with one at the $l$-th position and zeros elsewhere. By the definition of proposed encoding matrix, the obtained $\underline{s}_1$ is at the first row (systematic part) of $\Phi$. Thus, we have the equation $\underline{i}_{1}^{t}S=\underline{s}_1$ which will be utilized in the upcoming decoding stages.

In the $l$-th stage, $1\leq l\leq k$, the DC can access the $\{\underline{d}_{i_{l}}^{t}[l],\ldots ,\underline{d}_{i_{k}}^{t}[l]\}$ in the $l$-th column of lower triangular part of $D_{\mathrm{DC}}$, and the corresponding encoding rows are $\{\underline{\phi}_{i_{l}}^{t},\ldots ,\underline{\phi}_{i_{k}}^{t}\}$. In the previous stages, we obtain $l-1$ equations:
\[
\underline{i}_{j}^{t}\overline{s}_{l}=\underline{s}_j[l],\forall j=1,\ldots ,l-1.
\]
It is noted that the $\{\underline{i}_{j}^{t}|1\leq j\leq l-1\}$ are the first $l-1$ rows of $\Phi$. The above equations are combined to obtain
\begin{equation}
\begin{bmatrix}\underline{i}_{1}^{t}\\
\vdots\\
\underline{i}_{l-1}^{t}\\
\underline{\phi}_{i_{l}}^{t}\\
\vdots\\
\underline{\phi}_{i_{k}}^{t}
\end{bmatrix}\overline{s}_{l}=\begin{bmatrix}\underline{s}_1[l]\\
\vdots\\
\underline{s}_{l-1}[l]\\
\underline{d}_{i_{l}}^{t}[l]\\
\vdots\\
\underline{d}_{i_{k}}^{t}[l]
\end{bmatrix}.\label{eq:overlines_l}
\end{equation}
Let the $D_1^l$ denote the matrix at the left-hand-side of (\ref{eq:overlines_l}). To solve the $\overline{s}_l$ successfully, the $D_1^l$ should be non-singular. Then we have $\underline{s}_l=\overline{s}_l^t$, and the  $\underline{i}_{l}^{t}S=\underline{s}_{l}$ can be utilized in the upcoming decoding stages.

The non-singularity of $D_1^l$ is discussed below. In the $D_{1}^{l}$, the set $\underline{i}^{t}=\{\underline{i}_{1}^{t},\ldots ,\underline{i}_{l-1}^{t}\}$ are the first $l-1$ rows of $\Phi$, and the set $\underline{\phi}^t=\{\underline{\phi}_{i_{l}}^{t},\ldots ,\underline{\phi}_{i_{k}}^{t}\}$ are $k-l$ rows in $\Phi$. As any $k$ rows of $\Phi$ are non-singular, the $D_{1}^{l}$ is also non-singular, as long as the two sets are mutually exclusive $\underline{i}^{t}\cap \underline{\phi}^t=\emptyset$.  To satisfy this condition, the order of fragments $[g_{1},\ldots ,g_{k}]$ in $C_{\mathrm{DC}}$ should follow a special condition: For the systematic fragment $\underline{c}_{l}^{t}$, $1\leq l\leq k$, downloaded from the node $i_j$, the $\underline{c}_{l}^{t}$ is placed at the $g_{j}$-th row of $C_{\mathrm{DC}}$, where $g_{j}\leq l$.

\emph{Example 3:} By following the codes given by Example 1, we assume that the DC connects to nodes 1, 2, and 4 respectively corresponding to encoding rows $\begin{bmatrix}1 & 0 & 0 & 0\end{bmatrix}$, $\begin{bmatrix}0 & 1 & 0 & 0\end{bmatrix}$ and $\begin{bmatrix}1 & 4 & 3 & 6\end{bmatrix}$. The three rows of $C_{\mathrm{DC}}$ are arranged as 
\[
C_{\mathrm{DC}}=\begin{bmatrix}1 & 0 & 0 & 0\\
0 & 1 & 0 & 0\\
1 & 4 & 3 & 6
\end{bmatrix}M.
\]
The DC downloads the whole $C_{\mathrm{DC}}^{\Delta}$ and the lower triangular part of $C_{\mathrm{DC}}^{\Phi}$. The $C_{\mathrm{DC}}^{\Delta}$ possesses the equation given by
\[
C_{\mathrm{DC}}^{\Delta}=\begin{bmatrix}1 & 0 & 0\\
0 & 1 & 0\\
1 & 4 & 3
\end{bmatrix}T.
\]
By the equation, the $T$ can be solved to obtain $\{\tilde{u}_{4},\tilde{u}_{7},\tilde{u}_{9}\}$, where the tilde symbol $\tilde{\bullet}$ indicates the solved terms. By the solved $T$, the DC calculates the lower triangular part of $D_{\mathrm{DC}}$ via
\[
D_{\mathrm{DC}}=C_{\mathrm{DC}}^{\Phi}-\begin{bmatrix}0\\
0\\
6
\end{bmatrix}\begin{bmatrix}\tilde{u}_{4} & \tilde{u}_{7} & \tilde{u}_{9}\end{bmatrix}.
\]
Let $D[i,j]$ denote the entry of $D_{\mathrm{DC}}$ at the $i$-th row and $j$-th column. The accessible part of $D_{\mathrm{DC}}$ is
\[
\begin{bmatrix}D[1,1] & - & -\\
D[2,1] & D[2,2] & -\\
D[3,1] & D[3,2] & D[3,3]
\end{bmatrix}=\begin{bmatrix}1 & 0 & 0\\
0 & 1 & 0\\
1 & 4 & 3
\end{bmatrix}S,
\]
where the notation "$-$" indicates the inaccessible entries. Firstly, by the first column of $D_{\mathrm{DC}}$, the first column of $S$ is solved. The solved symbols possesses the equality:
\begin{equation}
\begin{bmatrix}\tilde{u}_{1} & \tilde{u}_{2} & \tilde{u}_{3}\end{bmatrix}=\begin{bmatrix}1 & 0 & 0\end{bmatrix}S.\label{eq:help1}
\end{equation}
Secondly, to decode the second column of $S$, we have
\[
\begin{bmatrix}\tilde{u}_{2}\\
D[2,2]\\
D[3,2]
\end{bmatrix}=\begin{bmatrix}1 & 0 & 0\\
0 & 1 & 0\\
1 & 4 & 3
\end{bmatrix}\begin{bmatrix}u_{2}\\
u_{5}\\
u_{6}
\end{bmatrix}.
\]
Then the symbols $\{u_{5},u_{6}\}$ are solved. The solved symbols possess the equality:
\begin{equation}
\begin{bmatrix}\tilde{u}_{2} & \tilde{u}_{5} & \tilde{u}_{6}\end{bmatrix}=\begin{bmatrix}0 & 1 & 0\end{bmatrix}S.\label{eq:help2}
\end{equation}
By the third column of $D_{\mathrm{DC}}$ and the (\ref{eq:help1})(\ref{eq:help2}), we have
\[
\begin{bmatrix}\tilde{u}_{3}\\
\tilde{u}_{6}\\
D[3,3]
\end{bmatrix}=\begin{bmatrix}1 & 0 & 0\\
0 & 1 & 0\\
1 & 4 & 3
\end{bmatrix}\begin{bmatrix}u_{3}\\
u_{6}\\
u_{8}
\end{bmatrix}.
\]
Then the symbol $u_{8}$ is solved successfully.

\subsubsection{Data collector downloads the upper triangular part of $C_{\mathrm{DC}}^{\Phi}$\label{sub:upper triangular C_DC}}
In this case, the DC accesses the upper triangular part of $D_{\mathrm{DC}}$ defined in (\ref{eq:D_DC-1}). The steps are very similar to the above decoding scheme. The decoding structure can be expressed as $k$ stages, and each stage extracts a column of $S$ in backward order. That is, the $l$-th stage extracts the $(k+1-l)$-th column $\overline{s}_{k+1-l}$ of $S$. In the $l$-th stage, the DC can access the $\{\underline{d}_{i_{1}}^{t}[l],\ldots ,\underline{d}_{i_{k+1-l}}^{t}[l]\}$ taken from the $(k+1-l)$-th column of $D_{\mathrm{DC}}$ in upper triangular part, and the corresponding encoding rows are $\{\underline{\phi}_{i_{1}}^{t},\ldots ,\underline{\phi}_{i_{k+1-l}}^{t}\}$. Furthermore, we also have $l-1$ equations by the previous stages: 
\[
\underline{i}_{j}^{t}\overline{s}_{l}=\underline{s}_j[l],\forall j=k+2-l,\ldots ,k.
\]
Those equations are combined to obtain
\begin{equation}
\begin{bmatrix}\underline{\phi}_{i_{1}}^{t}\\
\vdots\\
\underline{\phi}_{i_{k+1-l}}^{t}\\
\underline{i}_{k+2-l}^{t}\\
\vdots\\
\underline{i}_{k}^{t}
\end{bmatrix}\overline{s}_{k+1-l}=\begin{bmatrix}\underline{d}_{i_{1}}^{t}[l]\\
\vdots\\
\underline{d}_{i_{k+1-l}}^{t}[l]\\
\underline{s}_{k+2-l}[l]\\
\vdots\\
\underline{s}_{k}[l]
\end{bmatrix}.\label{eq:overlines_l-1}
\end{equation}
Let the $D_{2}^{l}$ denote the left-hand-side matrix in (\ref{eq:overlines_l-1}). To decode the $\overline{s}_{k+1-l}$, the $D_{2}^{l}$ should be non-singular, and this condition induces that $\{\underline{\phi}_{i_{1}}^{t},\ldots ,\underline{\phi}_{i_{k+1-l}}^{t}\}\cap\{\underline{i}_{k+2-l}^{t},\ldots ,\underline{i}_{k}^{t}\}=\emptyset$, for $1\leq l\leq k$. By the above condition, the systematic fragment $\underline{c}_{l}^{t}$ downloaded from the node $i_{j}$ is placed at the $g_{j}$-th row of $C_{\mathrm{DC}}$, where $1\leq l\leq g_{j}\leq k$. Then the $\overline{s}_{k+1-l}$ can be solved successfully, and the formula $\underline{i}_{k+1-l}^{t}S=\overline{s}_{k+1-l}^{t}$ is utilized in the upcoming decoding stages.

\subsubsection{The time-sharing policy to balance the bandwidth requirements on each
connected node}
In the above two partial downloading schemes, both partial downloading schemes have the disadvantage that the transmission amounts for $k$ connected nodes are excessively unbalanced. To overcome this drawback, we can iteratively switch the two partial downloading schemes during the whole transmission rounds. Specifically, if a node $i_j$ transmits the elements of a code fragment in the lower triangular of $C_{\mathrm{DC}}^{\Phi}$ at this transmission round, this node will transmit the elements of next code fragment in the upper triangular of $C_{\mathrm{DC}}^{\Phi}$ at the next transmission round. By this time-sharing policy, each node transmits $d-(k-1)/2$ symbols in each transmission round on average.

As stated in Sections \ref{sub:lower triangular C_DC} and \ref{sub:upper triangular C_DC}, the two partial downloading schemes respectively give two different conditions on the order $[g_{1},\ldots ,g_{k}]$ of the downloaded fragments in $C_{\mathrm{DC}}$. Since the time-sharing policy iteratively applies two partial downloading schemes, the two conditions should be satisfied simultaneously. The intersection of two conditions is that, the systematic fragment $\underline{c}_{l}^{t}$ downloaded from the node $i_{j}$ is placed at the $g_{j}$-th row of $C_{\mathrm{DC}}$, where $1\leq g_{j}=l\leq k$.

\section{Comparisons and discussions\label{sec:Comparisons-and-discussions}}
In this section, we compare the proposed codes with prior works. The results are briefly summarized in Tables \ref{tab:Comparisons-of-repair-by-transfe} and \ref{tab:Comparisons-of-Exact-MBR}. 

\subsection{Comparisons for Repair-by-transfer codes\label{sub:Comparisons-for-RBT}}

\begin{table}
\caption{\label{tab:Comparisons-of-repair-by-transfe}Comparisons for repair-by-transfer
codes over $GF(q)$.}
\begin{tabular}{c|c|c|c}
\hline 
 & Down. policy  & Range of $n$  & Enc. comp.\tabularnewline
\hline 
Shah et al. \cite{Shah}  & -  & ${n \choose 2}\leq q+1$  & $O(n^{4})$\tabularnewline
\hline 
Ours (Section \ref{sec:repair-by-transfer codes})  & Partial  & $n\leq q+1$  & $O(n^{3})$\tabularnewline
\hline 
\end{tabular}
\end{table}

\begin{figure}
\includegraphics[width=1\columnwidth]{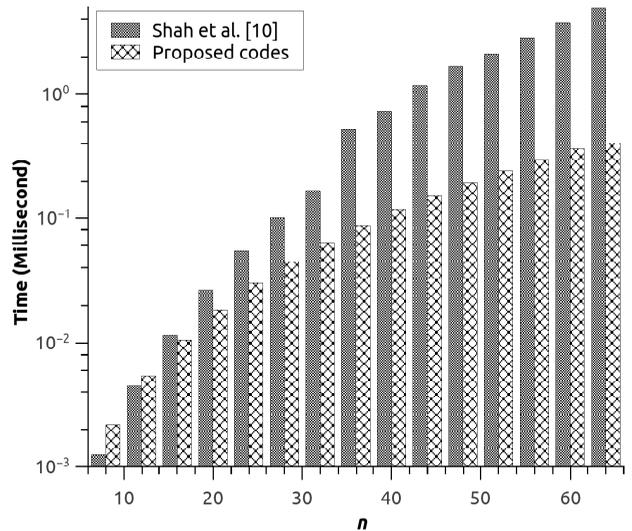}\caption{\label{fig:simulation}The simulations of \cite{Shah} and the proposed
repair-by-transfer codes.}
\end{figure}

This subsection compares the proposed repair-by-transfer codes with the \cite{Shah} introduced in Sec. \ref{sec:existing-repair-by-transfer}. As shown in Sec. \ref{sec:existing-repair-by-transfer}, the field size of \label{sub:Comparisons-for-RBT} is at least ${n \choose 2}\leq N+1$. For the proposed repair-by-transfer codes, Section \ref{sub:Coding-RBT} states that the feasible range of $n$ can be extended up to $n\leq N+1$ via the extended Vandermonde matrix. Hence we conclude that the size of finite field is significantly reduced.

Another issue is the computational complexities. We compare the complexities of both codes over the same finite field $GF(q)$. For the \cite{Shah}, it is evident that the $(N,K)=\left({n \choose 2},B\right)$ MDS code dominates the whole computational overhead. By employing the $\left({n \choose 2},B\right)$ systematic RS code, the encoding complexity is given by $O(({n \choose 2}-B)B)$. For the proposed scheme, the systematic version Sec. \ref{sub:Coding-RBT} computes the matrix $V$, whose computational cost is dominated by two terms $\check{\Phi}U_{\mathrm{R}}$ and the $\check{\Phi}U_{\mathrm{L}}\check{\Phi}^{t}$. As stated in Sec. \ref{sub:Coding-RBT}, both terms take a total of $O(k(n-k)^{2})+O(2k^{2}(n-k))=O(k(n^{2}-k^{2}))$ operations. To magnify the difference between both codes further, we consider the case $k=cn$ with a constant $c$. In this case, the big-O representation of both codes are simplified into $O(({n \choose 2}-B)B)=O(n^{4})$ , and $O(k(n^{2}-k^{2}))=O(n^{3})$, respectively. Thus, the proposed code reduces one order of magnitude in big-O complexity representation. The real simulations of two codes are shown in Fig. \ref{fig:simulation}. Both codes are written in JAVA, and the programs are running on Intel i7-950, 4GB RAM, Windows 8. We test the case $k=n/2$ at $n=\{8,12,\ldots ,64\}$ over $GF(2^{16})$. In the simulation, the source data are generated by a random number generator. The Y-axis represents the logarithm of the encoding time of the $B$ input symbols on average. As shown in Fig. \ref{fig:simulation}, the performance of the proposed codes is better than the \cite{Shah} if the $n$ is larger than a specific value. For the small value of $n$, we conjecture that the structure of \cite{Shah} is more simple, and the proposed algorithm contains a number of redundant arithmetic operations in the the congruence $V$, so that the \cite{Shah} is better.

\subsection{Comparisons for systematic Exact-MBR codes\label{sub:Compare-SystematicMBR}}
\begin{table*}
\centering
\caption{\label{tab:Comparisons-of-Exact-MBR}Comparisons for Exact-MBR codes
over $GF(q)$.}
\begin{tabular}{c|c|c|c|c}
\hline 
 & Syst. & Down. policy  & Range of $n$ & Enc. complexity\tabularnewline
\hline 
\multirow{2}{*}{Rashmi et al. \cite{Rashmi}} & N & Full & $n\leq q$ & $O(nd^2)$\tabularnewline
\cline{2-5} & Y & Full & $n\leq (q+k-d)$ or $n\leq q$ & $O(nd^2)$\tabularnewline
\hline 
Gong and Wang\cite{Gong} & N & Partial & $n\leq q$ & $O(nd^2)$\tabularnewline
\hline 
Ours (Section \ref{sec:Systematic-Exact-MBR}) & Y & Partial & $n\leq q$ & $O(nd^2)$ or $O(n\log n)$\tabularnewline
\hline 
\end{tabular}
\end{table*}
In the following, we compare the proposed systematic Exact-MBR codes with the \cite{Rashmi}, in terms of the range of $n$ and the encoding complexity. For the range of $n$, the \cite{Rashmi} presents two distinct forms for the encoding matrix, so the $n$ has two distinct upper bounds. The first form is expressed as
\begin{equation}
\Psi=\left[\begin{array}{cc}
I_{k} & \mathbf{0}\\
\tilde{\Phi} & \tilde{\Delta}
\end{array}\right],\label{eq:Rashmi's matrix}
\end{equation}
where $I_{k}$ denotes a $k\times k$ identity matrix, $\mathbf{0}$ is a $k\times(d-k)$ zero matrix. The $[\begin{array}{cc} \tilde{\Phi} & \tilde{\Delta}\end{array}]$ is a $(n-k)\times d$ Cauchy matrix, where the sizes of $\tilde{\Phi}$ and $\tilde{\Delta}$ are $(n-k)\times k$ and $(n-k)\times(d-k)$, respectively. As stated by \cite{Rashmi}, the (\ref{eq:Rashmi's matrix}) meets the two conditions of Exact-MBR encoding matrix. By definition, a $(n-k)\times d$ Cauchy matrix requires $n-k+d$ distinct symbols. As the $GF(q)$ contains a total of $q$ distinct symbols, the feasible range of $n$ is 
\[
n-k+d\leq q\Rightarrow n\leq q+k-d.
\]
As addressed in Sec. \ref{sub:PSRS}, the range of $n$ for the PSRS codes is $n\leq q$, so is the proposed Exact-MBR code. Due to $k\leq d$, the proposed codes have larger range of $n$.

In the second form of encoding matrix \cite{Rashmi}, the range of $n$ is also $n\leq q$. However, the second form is not explicit and the matrix generation requires an additional matrix inversion and multiplication step. An explicit form can facilitate the further development on the codes. For example, the partial decoding algorithm proposed in Sec. \ref{sub:DataRecAlg-Min} is based on the observations on the form of encoding matrix. If the encoding matrix is not explicit, the partial decoding algorithm may become more difficult to be designed. Furthermore, by Appendix, the proposed $(n, k)$ PSRS codes can be implemented by generator polynomials. The size of generator polynomial is $(n-k)$, which is lower than the size of encoding matrix $(n-k)\times k$ in parity part. Thus, the generator polynomial approach is more common in usage.

The encoding complexities of those codes are discussed below. Suppose those three codes are implemented with native matrix product approach. As the sizes of encoding matrix and message matrix are $n\times d$ and $d\times d$ for the three codes, the encoding complexity is $O(nd^2)$. Furthermore, Section \ref{sub:PSRS} indicates that the PSRS codes can be implemented with fast Fourier transforms. By FFTs, the encoding complexity can be reduced to $O(n\log n)$.

\subsection{Comparisons for partial downloading schemes on Exact-MBR codes\label{sub:Compare-PDS}}
The partial downloading scheme is useful to reduce the requisite throughput to reconstruct data. This subsection highlights the differences between \cite{Gong} and ours. First, the proposed scheme requires that the systematic fragments should be placed at a specific row of $C_\mathrm{DC}$. On the other hand, the \cite{Gong} do not require this condition as the \cite{Gong} is developed on non-systematic codes. Second, in our survey, this is the first work of considering the throughput balance on the connected nodes.

\subsection{Simulations for systematic Exact-MBR codes\label{sub:Fast-SystematicMBR}}
\begin{figure}
\includegraphics[width=1\columnwidth]{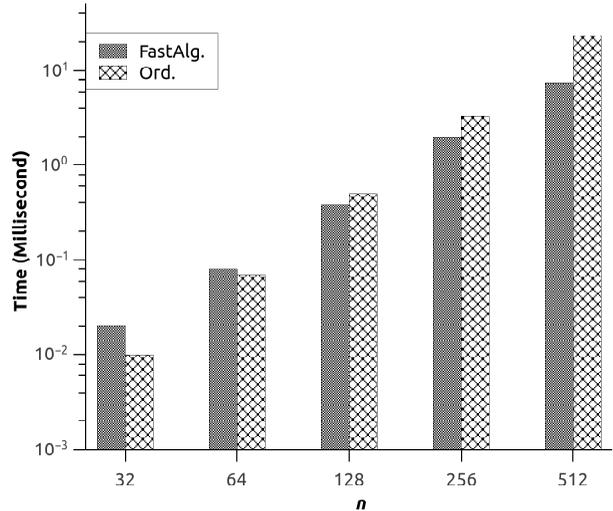}\caption{\label{fig:simulation2}The simulations of systematic Exact-MBR codes with native approach and fast approach.}
\end{figure}
As stated in Section \ref{sub:PSRS}, the PSRS codes can be implemented with FFT. By employing the fast algorithm of PSRS code, we expected that the encoding time of Exact-MBR codes can be reduced. Based on this motivation, we implement the native and fast approaches of Exact-MBR codes, and the simulation results are shown in Figure \ref{fig:simulation2}. Both codes are written in JAVA, and the programs are running on Intel i7-950, 4GB RAM, Windows 8. We test the case $k=3/8\times n$, $d=n/2$, at $n\in\{32, 64, 128, 256, 512\}$ over Fermat field $GF(2^{16}+1)$. As shown in Figure \ref{fig:simulation2}, the fast approach works better for larger $n$. Otherwise, the native approach is suggested.

\section{Conclusions\label{sec:Conclusions}}
The contributions of this paper can be organized in two parts. First, a new class of repair-by-transfer codes are proposed at $d=n-1$ MBR points. As compared with prior works, the proposed repair-by-transfer code demands smaller finite field and lower big-O complexity. The partial downloading scheme is also developed on the proposed repair-by-transfer codes to avoid the unnecessary symbol transmissions. The simulation shows that the proposed repair-by-transfer codes require fewer arithmetic operations than the prior work when $n$ is larger than a specific value. Second, for all feasible parameters $(n,k,d)$, we present an encoding matrix for systematic Exact-MBR codes via the partially systematic Reed-Solomon codes. To minimize the number of transmitted symbols in data reconstruction process, the partial downloading scheme is designed on the proposed Exact-MBR codes. However, the transmission amount for those connected nodes are excessively unbalanced. Thus, a time-sharing scheme is presented to balance the bandwidths requirements on those connected nodes. The proposed Exact-MBR codes can be implemented via fast Fourier transforms. As shown in the simulations, the fast approach has better encoding performance for large $n$.

\appendices
\section{Partially systematic Reed-Solomon codes by generator polynomial\label{sec:Partially-RS}}
The appendix presents another approach of $(n,k,d)$ PSRS codes by generator polynomials. In this approach, the messages and codewords are formulated as polynomials. Thus, the message $\underline{a}$ and $\underline{b}$ are
\[
a(x)=a_{0}+a_{1}x+\ldots+a_{k-1}x^{k-1};
\]
\[
b(x)=b_{0}+b_{1}x+\ldots+b_{d-k-1}x^{d-k-1}.
\]
The codeword polynomial is defined as
\[
c(x)=c_{0}(x)+c_{1}(x),
\]
where the $c_{0}(x)$ is the codeword generated from $a(x)$ via $(n,k)$ systematic RS code, and the $c_{1}(x)$ is the codeword generated from $b(x)$ via $(n-k,d-k)$ RS code. Precisely, for the construction of $c_{0}(x)$, the generator polynomial of $(n,k)$ systematic RS code is defined as
\[
g_{0}(x)=(x-1)(x-\alpha)\ldots(x-\alpha^{n-k-1}).
\]
Then the parity polynomial $r_{0}(x)$ is calculated through polynomial division
\[
r_{0}(x)=x^{n-k}a(x)\textrm{ (mod }g_{0}(x)).
\]
The codeword $c_{0}(x)$ is expressed as the concatenation of $a(x)$ and $r_{0}(x)$:
\begin{equation}
c_{0}(x)=x^{n-k}a(x)-r_{0}(x).\label{eq:c0(x)}
\end{equation}
For the construction of $c_{1}(x)$, the generator polynomial of $(n-k,d-k)$ RS code is defined as
\[
g_{1}(x)=(x-1)(x-\alpha)\ldots(x-\alpha^{n-d-1}).
\]
The $c_{1}(x)$ can be formed as the systematic or non-systematic version. For the systematic case, the codeword polynomial is defined as
\[
r_{1}(x)=x^{n-d}b(x)\textrm{ (mod }g_{1}(x));
\]
\begin{equation}
c_{1}(x)=x^{n-d}b(x)-r_{1}(x).\label{eq:c1(x)}
\end{equation}
The polynomial $a(x)$ is embedded in the $c(x)$ between $x^{n-k}$ and $x^{n-1}$, as the degree of $c_{1}(x)$ is less than $n-k$. Thus, the partially systematic condition holds. By generator polynomial, the length of this $(n,k,d)$ coding algorithm gets up to $n\leq q-1$ over $GF(q)$. The decoding algorithms are explained in the following.

\subsection{Full erasure decoding from $d$ codeword symbols}
The $a(x)$ and $b(x)$ can be reconstructed by arbitrary $d$ out of $n$ coefficients of the $c(x)$. As $g_{0}(x)$ and $g_{1}(x)$ are respectively the factors of $c_{0}(x)$ and $c_{1}(x)$, the $\gcd(g_{0}(x),g_{1}(x))=g_{0}(x)$ is also the factor of the $c(x)$. Therefore the $(n,k,d)$ PSRS code is isomorphic to the $(n,d)$ RS code with the generator polynomial $g_{0}(x)$. Thus, the $c(x)$ can be reconstructed from arbitrary $d$ out of $n$ coefficients via Forney algorithm. Forney algorithm is a method to compute the erasures of BCH codes at known error locations. When the $c(x)$ is completely reconstructed, the $a(x)$ is located in the systematic part of $c(x)$. Then the $c_0(x)$ can be computed from $a(x)$, and subsequently the $c_{1}(x)=c(x)-c_{0}(x)$. Thus, the $b(x)$ is decoded from $c_{1}(x)$.

\subsection{Partial erasure decoding from $k$ codeword symbols}
Given the $b(x)$, the message $a(x)$ can be reconstructed by arbitrary $k$ out of $n$ coefficients in $c(x)$. By (\ref{eq:c1(x)}), the $c_{1}(x)$ is calculated from $b(x)$. Since we have $k$ coefficients in $c(x)$, the corresponding $k$ coefficients in $c_{0}(x)=c(x)-c_{1}(x)$ can also be calculated. As the $c_{0}(x)$ is the codeword of $(n,k)$ systematic RS code, the $c_{0}(x)$ can be completely recovered via Forney algorithm. Then the message $a(x)$ is obtained from $c_{0}(x)$.

\begin{IEEEbiography}
[{\includegraphics[width=1in]{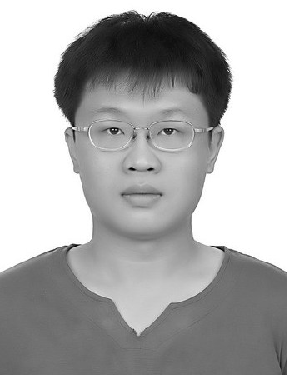}}]{Sian-Jheng
Lin}was born in Taichung, Taiwan, in 1981. He received the B.S.,
M.S., and Ph.D. degrees in computer science from National Chiao Tung
University, in 2004, 2006, and 2010, respectively. He is currently
a postdoctoral fellow with the Research Center for Information Technology
Innovation, Academia Sinica. His recent research interests include
data hiding and error control coding.
\end{IEEEbiography}
\begin{IEEEbiography}
[{\includegraphics[width=1in]{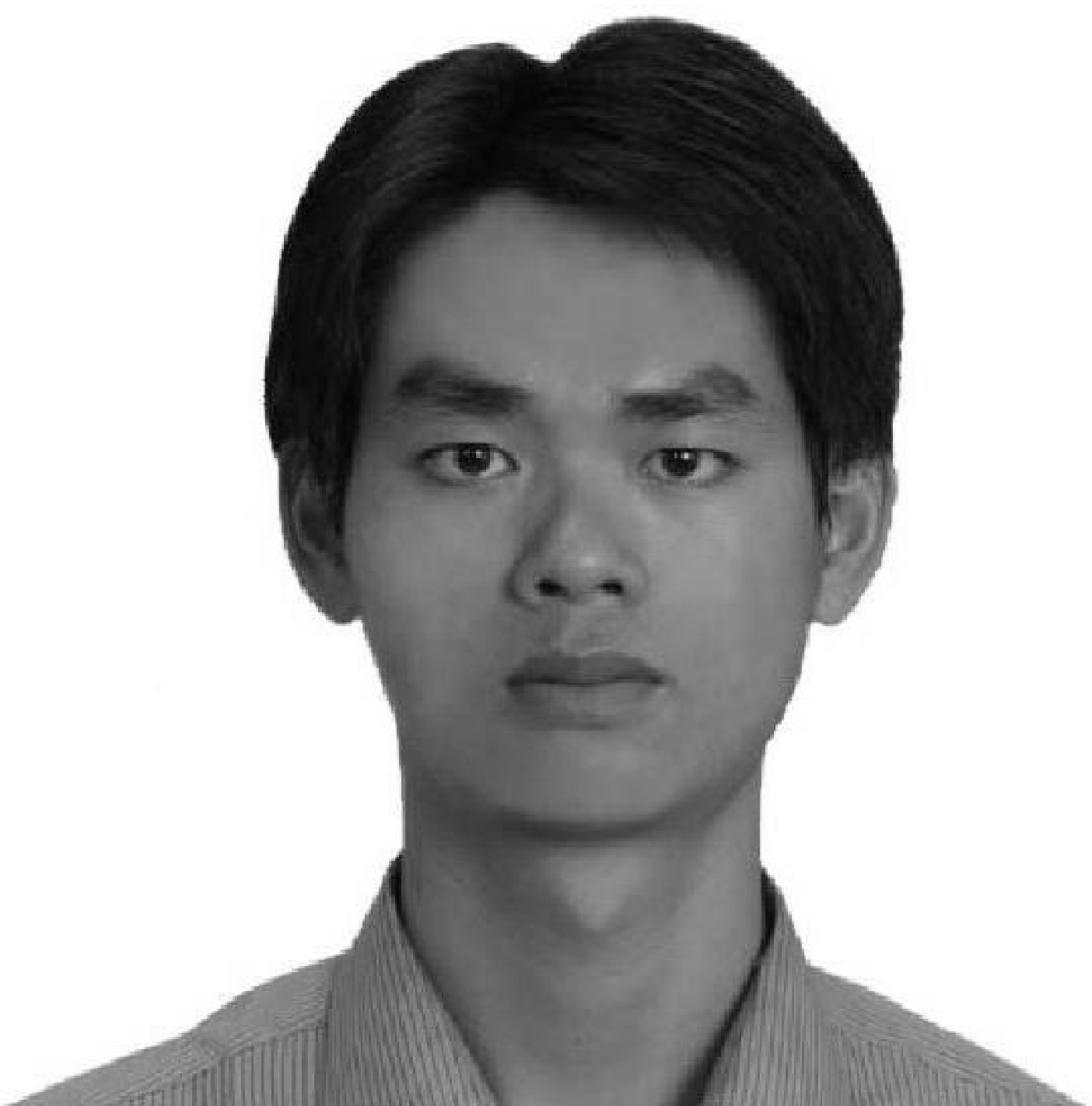}}]{Wei-Ho Chung}
was born in Kaohsiung, Taiwan, in 1978. He received the B.Sc. and
M.Sc. degrees in Electrical Engineering from National Taiwan University,
Taipei City, Taiwan, in 2000 and 2002 respectively. From 2005 to 2009,
he was with the Electrical Engineering Department at University of
California, Los Angeles, where he obtained his Ph.D. degree. From
2000 to 2002, he worked on routing protocols in the mobile ad hoc
networks in the M.Sc. program in National Taiwan University. From
2002 to 2005, he was a system engineer at ChungHwa Telecommunications
Company, where he worked on data networks. In 2008, he was an research
intern working on CDMA systems in Qualcomm, Inc. From 2007 to 2009,
he was a Teaching Assistant at UCLA. From June to December 2009, Dr.
Chung had been working as a research associate in San Diego, California,
on wireless communications for multimedia communications and unequal
error protection for video transmission. His research interests include
communications, signal processing, and networks. Dr. Chung received
the Taiwan Merit Scholarship from 2005 to 2009, and the Best Paper
Award in IEEE WCNC 2012. Dr. Chung has been an assistant research
fellow in Research Center for Information Technology Innovation in
Academia Sinica, Taiwan, since January 2010.
\end{IEEEbiography}
\end{document}